\documentclass[aps,nofootinbib,superscriptaddress, showpacs,preprintnumbers,  nofootinbibt,twocolumn]{revtex4-1}
\usepackage{eurosym}
\usepackage{amsmath}
\usepackage{bm}
\usepackage{amsfonts}
\usepackage{amssymb}
\usepackage{graphicx}
\usepackage{hyperref}
\usepackage{mathtools}
\usepackage{color}
\usepackage{physics}

\usepackage{graphicx}
\usepackage{dcolumn}
\usepackage{bm}
\usepackage{caption}
\usepackage{subcaption}
\usepackage{epstopdf}
\usepackage{amsmath}
\usepackage{xcolor}

\def\be{\begin{equation}}
\def\ee{\end{equation}}
\def\bea{\begin{eqnarray}}
\def\eea{\end{eqnarray}}

\begin{document}

\title{Thermodynamics of the Weyl Geometric Gravity Black Holes}
\author{Muhammad F. A. R. Sakti}
\email{fitrahalfian@gmail.com}
\affiliation{High Energy Physics Theory Group, Department of Physics,
Faculty of Science, Chulalongkorn University, Bangkok 10330, Thailand,}
\author{Piyabut Burikham}
\email{piyabut@gmail.com}
\affiliation{High Energy Physics Theory Group, Department of Physics,
Faculty of Science, Chulalongkorn University, Bangkok 10330, Thailand,}
\author{Tiberiu Harko}
\email{tiberiu.harko@aira.astro.ro}
\affiliation{Department of Theoretical Physics, National Institute of Physics
and Nuclear Engineering (IFIN-HH), Bucharest, 077125 Romania,}
\affiliation{Department of Physics, Babes-Bolyai University, Kogalniceanu Street,
	Cluj-Napoca 400084, Romania,}
\affiliation{Astronomical Observatory, 19 Ciresilor Street,
	Cluj-Napoca 400487, Romania,}

\begin{abstract}
We consider the thermodynamic properties of an exact black hole solution obtained in Weyl geometric gravity theory, by considering the simplest conformally invariant action, constructed from the square of the Weyl scalar, and the strength of the Weyl vector only. The action is linearized in the Weyl scalar by introducing an auxiliary scalar field, and thus it can be reformulated as a scalar-vector-tensor theory in a Riemann space, in the presence of a nonminimal coupling between the Ricci scalar and the scalar field. In static spherical symmetry, this theory admits an exact black hole solution, which generalizes the standard Schwarzschild-de Sitter solution through the presence of two new terms in the metric, having a linear and a quadratic dependence on the radial coordinate, respectively. The solution is obtained by assuming that the Weyl vector has only a radial component. After studying the locations of the event and cosmological horizons of the Weyl geometric black hole, we investigate in detail the thermodynamical (quantum properties) of this type of black holes, by considering the Hawking temperature, the entropy,  specific heat and the Helmholtz free energy functions on both the event and the cosmological horizons. The Weyl geometric black holes have thermodynamic properties that clearly differentiate them from similar solutions of other modified gravity theories.  The obtained results may lead to the possibility of a better understanding of the properties of the black holes in alternative gravity, and of the relevance of the thermodynamic aspects in black hole physics.
\end{abstract}

\date{\today }
\maketitle

\section{Introduction}

Einstein's general theory was historically the first approach that succeeded in giving a full description of gravitational phenomena in geometric terms
\cite{Einstein:1915}. The development of general relativity took place within the framework of Riemannian geometry \cite{Riem}, which provided a powerful mathematical instrument necessary for the understanding of the gravitational phenomena. Extremely important in the evolution of gravitational theories was the variational derivation of the gravitational field equations, which was realized  by Hilbert \cite{Hilbert:1915}.  General relativity gives a remarkably good description of the gravitational effects in the Solar System, like, for example,  the perihelion precession of the planets, the bending of light by the Sun, or the Shapiro time delay effect \cite{Will, Marchi}. One of the most important, and at the same time intriguing, predictions of general relativity is the existence of gravitational waves,  which were detected recently \cite{GW1,GW2, AgazieNanoGrav2023, AgazieNanoGrav2023L9, AfzalNanoGrav2023}.

However, despite these important achievements, on both small and large scales, the theory of general relativity  faces several important challenges.  The precise astronomical observations of the Type IA Supernovae led to an unexpected discovery, namely, that the present-day Universe is in a state of accelerated, de Sitter type expansion. The accelerated expansion  is assumed to be caused by a special form of matter, called dark energy \cite{R1,Per}. The cosmological constant $\Lambda$ introduced by  Einstein in the field equations \cite{Ein1} is a particular, and the simplest form of such a fluid. However, the physical, or geometrical natures of the cosmological constant, and of the dark energy are not yet known \cite{Wein, AmendolaTsujikawa2010, JoycePhysRep2015, JoyceAnnuRev2016, TawfikDahab2019, FrusciantePhysRep2020}.  Another unsolved problem, which is also a fundamental one, is the dark matter problem. This problem has its origins in the surprising dynamics of the galactic rotation curves \cite{Oks}. General relativity cannot describe the behavior of hydrogen clouds gravitating outside the baryonic matter distribution of the galaxies without postulating the existence of an exotic form of matter, called dark matter \cite{Bian}.

The $\Lambda$CDM standard cosmological paradigm, based on general relativity with the inclusion of a cosmological constant, is also confronted with several significant challenges. An intriguing open problem in present day cosmology is related to the deviations between the Hubble expansion rates as obtained from the Cosmic Microwave Background (CMB) experiments, and the small redshift determinations, based on the studies of the Type IA supernovae. These dissimilarities in the values of the Hubble constant $H_0$ for $z=0$ represent the so called Hubble tension \cite{Val}.

If the Hubble tension does indeed exist, and it is not only an observational artefact, it strongly points towards the requirement of considering, and analyzing in depth alternative theories of gravity. These new theories must go beyond standard general relativity,  and they may provide new and unexpected solutions to the dark energy and dark matter problems.

Almost immediately after the correct field equations of general relativity were written down, Schwarzschild obtained the first vacuum solution of the field equations \cite{Sch}, for a static, point-like mass distribution. This solution led to the introduction of a new astrophysical concept, and of a new type of objects, called black holes. The external solution of the Einstein equations for a spinning object was obtained around sixty years later by Kerr \cite{Kerr}.  Black holes are essentially  very simple objects, characterized by the presence of an event horizon, and of a central singularity. An important result in astrophysics states the existence of a limiting mass for neutron stars of the order of $3.2M_{\odot}$ \cite{Ruf}.  Recently, the Event Horizon Telescope collaboration presented
the first images of the supermassive black hole M87* \cite{EH1,EH2,EH3}. These observations seem to point towards a Kerr-like nature for the Sgr A* black hole. However, based on these observations, one cannot explicitly eliminate the possibility of the existence of deviations with respect to the predictions of general relativity.

The Schwarzschild and the Kerr solutions raised some fundamental, and essentially still unsolved problems, related to the existence of singularities in black hole solutions \cite{SH1,SH2}.

  Three years after the final form of the gravitational field equations was found, Weyl \cite{Weyl1, Weyl2} did introduce a generalization of the Riemannian geometry. Weyl's  main goal was to formulate a unified theory of gravity and electromagnetism. In vacuum the Maxwell equations of electromagnetism are conformally invariant. Weyl suggested that the gravitational field equations  must have the same symmetry. The conformal invariance of the physical laws was introduced in a general way by Weyl through the creation of a new geometry.  Weyl's geometry is nonmetric, with the covariant derivative of the metric tensor not vanishing identically, and thus $\nabla _{\mu}g_{\alpha \beta}=Q_{\mu \alpha \beta}=\omega_{\mu}g_{\alpha \beta}$, with $Q_{\mu \alpha \beta}$ representing the nonmetricity tensor, and $\omega _{\mu}$ is the Weyl vector field. In his original theory Weyl assumed that $\omega _\mu$ is the electromagnetic four-potential $A_\mu$.  In Weyl's geometry, the length of vectors change during their parallel transport from the point $x^\mu$ to $x^\mu+dx^\mu$. This property led Einstein to strongly criticise the  physical interpretation of Weyl's theory. For a presentation of the historical development of Weyl's geometry,  and of its  recent physical applications, see \cite{Weyl3}.

However, if one abandons the interpretation of the Weyl vector field as corresponding to the electromagnetic potential $A_\mu$,  Weyl's geometry becomes a very interesting  generalization of Riemann's geometry, with an important potential for cosmological and gravitational applications.

A generalization of  Weyl's theory was proposed by Dirac \cite{Di1,Di2}, who extended the original theory of Weyl by introducing a real scalar field $\beta$, of weight $w(\beta)=-1$.  The cosmological applications of  Dirac's theory were considered in detail in \cite{Di3}, \cite{Di4}, and \cite{Isrcosm}, respectively. Other generalizations of the Weyl theory, and of its physical applications, have been discussed in \cite{Ut1,Ut2,Ni}.

 A conformally invariant gravitational theory, formulated in Riemannian geometry, with action $S_{Weyl} =-\frac{1}{4}\int d^4x\sqrt{-g}
C_{\mu\nu\rho\sigma}C^{\mu\nu\rho\sigma}$,
where $C_{\mu\nu\rho\sigma}$ is the Weyl tensor, was proposed and thoroughly investigated in \cite{M1,M2,M3,M4,M5,M6}. The theories with actions constructed by using the Weyl tensor  are called conformally invariant, or Weyl type gravity theories.  
An exact vacuum solution  of the Weyl gravity theory, given by
 $
 A(r)=1-3\beta \gamma -\frac{\beta\left(2-3\beta \gamma\right)}{r}+\gamma r+kr^2,
 $
 where $\beta$, $\gamma$ and $k$ are constants, was found in \cite{M1}. This solution may provide an alternative solution of the dark matter problem, since it can explain the observational data without introducing a dark matter component \cite{M1}. A metric similar in form to the exact Weyl gravity vacuum solution was found in \cite{Pi} as a solution of the field equations of the dRGT massive gravity theory.

 The important role the conformal transformations may play in cosmology was considered by Penrose \cite{P1}, who introduced a cosmological model called Conformal Cyclic Cosmology (CCC). The CCC theory assumes that during both the de Sitter era, and the Big Bang, the spacetime is conformally flat. The cosmological and physical implications of the CCC model were investigated in \cite{P1a, P2,P3,P4,P5,P6,P7}.  In \cite{H1} 't Hooft proposed to explain the small-scale characteristics of gravity by a breaking of the conformal symmetry invariance.  In \cite{H2} a theory of gravity was introduced,  in which  the conformal component of the metric is interpreted as a dilaton field, with black holes being regular solitons, topologically trivial, without any horizons, firewalls, or singularities.

Weyl geometry is the mathematical basis of the $f(Q)$ gravity theory, also called the symmetric teleparallel gravity, first introduced in \cite{Q1}. In this approach  it is assumed that the nonmetricity $Q$ is the fundamental  geometrical quantity that describes gravity. The initial analysis of \cite{Q1} was generalized in \cite{Q2}, and led to the creation of the $f(Q)$ gravity theory, sometimes also called the nonmetric gravity theory. In the $f(Q)$ theory, the action is defined as $S=\int{\left(f(Q)+L_m\right)\sqrt{-g}d^4x}$, where $f$ is an arbitrary function of the nonmetricity. The cosmological and astrophysical implications of the modified $f(Q)$ gravity theory have been extensively investigated in \cite{Q2, Q3, Q4, Q5, Q6}.

A novel point of view on the Weyl geometry, and on its applications in gravity, was introduced, and analyzed in \cite{Gh1,Gh2,Gh3,Gh4,Gh5,Gh6,Gh7, Gh8,Gh9,Gh10, Gh11}.  As a starting point, the simplest possible gravitational action in Weyl geometry was adopted, constructed additively from the square of the scalar Weyl curvature $\tilde{R}^2$, and the square of the strength of the Weyl vector field $\tilde{F}_{\mu \nu}^2$. The important new characteristic of the theory is the linearization of the quadratic term in the Weyl scalar, by introducing an auxiliary scalar field. Thus, the vector-tensor Weyl theory can be reformulated as a scalar-vector tensor theory, linear in the Weyl, and the Ricci scalars.  The quadratic Weyl geometric action has a spontaneous symmetry breaking through a Stueckelberg mechanism, with the Weyl gauge field becoming massive. Hence, the scalar field is described by a Proca like action \cite{Gh1, Gh2,Gh3}. Through this symmetry breaking mechanism the Weyl geometric action recovers the Einstein-Hilbert action of general relativity, together with a positive cosmological constant of geometric nature \cite{Gh1,Gh2,Gh3}.

Conformally invariant couplings of geometry and matter in Weyl geometric gravity were analyzed in \cite{HT1}, leading to the formulation of the conformally invariant $f\left(R,L_m\right)$ theory of gravity \cite{HT1}. The conformally invariant $f\left(R,L_m\right)$ theory accurately describes the cosmological data for the Hubble function up to a redshift of around $z\approx 3$.  The conformally invariant $f\left(R,L_m\right)$ theory of gravity was studied in the Palatini formulation in \cite{HT2}.

In the Weyl geometric gravity theory black hole type solutions in spherical symmetry were investigated in detail in \cite{HT3}, by using numerical and analytical methods. The quantum thermodynamic properties of the obtained classes of Weyl geometric black holes were also considered. Moreover, an exact static spherically symmetric three parametric black hole solution  was obtained, which represents an extension of the Schwarzschild - de Sitter solution of general relativity, in the presence of a scalar field, and of the Weyl vector.

The exact solution of the Weyl geometric gravity was applied for the description of the properties of the galactic rotation curves in \cite{HT4}.  A comparison of the predictions of the exact Weyl geometric gravity solution with a small sample of galactic rotation curves was also performed. To take into account the effects of the baryonic matter the existence of an explicit breaking of the conformal invariance at the galactic level was assumed. An extensive comparison between the predictions of the Weyl geometric dark matter model and the observational data of the SPARC database was performed in \cite{HT5}, with the results confirming the possibility of a Weyl geometric interpretation of the dynamics of galaxies as proposed in \cite{HT4}. The properties of the compact objects in Weyl geometric gravity were investigated in \cite{HT6}.

Black holes have many fascinating properties, but perhaps the most interesting ones are related to their thermodynamical behaviors. The black hole thermodynamics is constructed through a subtle relationship uniting general gravity, quantum field theory, thermodynamics, and statistical mechanics. It
gives a powerful insight into the  nature and the properties of the black holes, and can lead to the understanding of the possible profound relation existing between  quantum theory and gravitational physics. The investigations of the black hole thermodynamics did begin with the study of Hawking \cite{Haw1}, in which it was proved that the area of the event horizon of a black hole does not decrease \cite{Haw1}. This important result is known as the area theorem. Bekenstein introduced the concept of entropy, which is proportional to the area of the black hole’s horizon \cite{Haw2, Haw3}.
The four laws of black hole thermodynamics were established in \cite{Haw4}, and they have a form similar to the four laws of standard thermodynamics.

An important result in black hole physics was the suggestion of the existence of the Hawking radiation \cite{Haw5}, a result which follows from the application of the quantum field theory in curved spacetimes. Moreover, this analysis also allows to establish the relation between the temperature of a black hole and its surface gravity. For detailed reviews of black hole thermodynamics see \cite{Dav,Wald1,Wald2}. For some recent investigations on the thermodynamical or astrophysical properties of black holes see  \cite{Piel, Bambi}, and \cite{Xu}, respectively.

Hence, in the physics and astrophysics of black holes the study of their thermodynamic properties is one of the most interesting subjects to study. It is the goal of the present paper to consider the thermodynamical properties of the Weyl geometric black holes \cite{HT3,HT4}, obtained as solutions of the vacuum field equations of Weyl geometric gravity. In order to understand the thermodynamic properties one needs first to analyze the horizons of the black hole solution. It turns out that the Weyl geometric black hole can have two event horizons, and one cosmological horizon, respectively. The mass of the black is also an important indicator of its astrophysical and thermodynamical properties. The Hawking temperature of the Weyl geometric black hole is obtained in terms of its surface gravity. The entropies of the event and of the cosmological horizons are fundamental quantities that allows the understanding of the thermodynamic properties of the black holes. To obtain the entropy of the Weyl geometric black hole we use the Noether charge approach. The stability of the black hole is investigated by considering the behavior of the Helmholtz and Gibbs free energy. The heat capacity of the Weyl geometric black hole is also investigated in detail. The first law of thermodynamics on the cosmological horizon for the Weyl geometric black hole is also formulated. In each case the dependence of the thermodynamical quantities on the parameters of the solution are fully investigated.

The present paper is organized as follows. We review the basics of the Weyl geometric gravity in Section~\ref{sect1}, where the vacuum field equations in static spherical symmetry are also written down.  The exact solution of the field equations is introduced in Section~\ref{sect2}, where the position of the horizons, and the effective energy density and pressure associated to the solution are discussed. A possible relation with the dark matter problem is also considered.  The thermodynamic properties of the Weyl geometric black hole on the event horizon are investigated in Section~\ref{sect3}. The black hole mass, Hawking temperature, entropy and volume are considered in detail. The thermodynamic stability of the Weyl geometric black hole is analyzed in Section~\ref{sect4}, by considering the heat capacity and the Helmholtz and Gibbs energies, respectively. The Hawking luminosity and the evaporation time of the Weyl geometric black holes are estimated in Sect.~\ref{sec:evap}. The thermodynamics of the cosmological horizon of the Weyl geometric black hole is analyzed in Section~\ref{sect5}.  We discuss and conclude our results in Section~\ref{sect6}. The fundamentals of the Weyl geometry, necessary for the understanding of Weyl geometric gravity, are presented in Appendix\ref{app:1}.

\section{Weyl Geometric Gravity}\label{sect1}

In the present Section we will briefly review the theoretical formalism of the Weyl geometric gravity, and we present the exact vacuum solution of this theory. We also reformulate the static, spherically symmetric field equations in terms of an effective density, and effective pressure. For further details on the mathematical and physical aspects of Weyl geometric gravity we refer the readers to \cite{HT4}.

\subsection{Field equations}

The Weyl geometric gravity theory is represented by the following action \cite{HT4}
\begin{eqnarray}\label{eq:actionWeyl}
	\mathcal{S} &=& \int d^4x\sqrt{-g} \bigg[ -\frac{1}{12}\frac{\phi^2}{\xi^2}\left(R -3\alpha\nabla_\mu\omega^\mu -\frac{3}{2}\alpha^2\omega_\mu\omega^\mu \right) \nonumber\\
	& & -\frac{1}{4}\frac{\phi^4}{\xi^2}-\frac{1}{4}F_{\mu\nu}F^{\mu\nu} \bigg].\
\end{eqnarray}

This action has its physical origins in Weyl geometry, and it is obtained through the linearization, with be help of an auxiliary scalar field, of the action $S=\int{\left(\tilde{R}^2-\tilde{F}_{\mu \nu}\tilde{F}^{\mu \nu}\right)\sqrt{-g}d^4x}$, where $\tilde{R}$ is the Weyl scalar, defined in the Weyl geometry, and $\tilde{F}_{\mu \nu}$ is the strength of the Weyl vector $\omega _\mu$. By $\phi$ we have denoted the scalar degree of freedom of the theory.  For a brief review of the basic concepts of Weyl geometry and of Weyl geometric gravity see Appendix~\ref{app:1}.

Note that the effective gravitational constant in this theory is $G = -3\xi^2/(4\pi \phi^{2})$. To omit the non-minimal coupling between gravity and scalar field, one can work in the Einstein frame rather than in this Jordan frame. However, in the present work we will work in the Jordan frame only.

By varying the action (\ref{eq:actionWeyl}) with respect to the metric tensor, we find the gravitational field equation below
\begin{eqnarray}\label{eq:GravEq}
&&\frac{\phi^2}{\xi^2}\left(R_{\mu\nu}-\frac{1}{2}Rg_{\mu\nu}\right)-\frac{1}{4\xi^2}\phi^4g_{\mu\nu}+\frac{1}{\xi^2}\left(g_{\mu\nu}\Box-\nabla_\mu \nabla_\nu \right)\phi^2 \nonumber\\
&&-\frac{3\alpha}{2\xi^2}\left(\omega^\rho\nabla_\rho \phi^2g_{\mu\nu}-\omega_\nu\nabla_\nu\phi^2-\omega_\mu\nabla_\nu\phi^2 \right)+6F_{\rho\mu}F_{\sigma\nu}g^{\rho\sigma}\nonumber\\
&&-\frac{3}{2}F^2_{\rho\sigma}g_{\mu\nu}+\frac{3\alpha^2}{4\xi^2}\phi^2\left(\omega_\rho\omega^\rho g_{\mu\nu}-2\omega_\mu\omega_\nu \right) =0.
\end{eqnarray}

By taking the trace of Eq.~(\ref{eq:GravEq}) and defining $\Phi =\phi^2$, we find
\begin{equation}
	\Phi R + 3\alpha\omega^\rho\nabla_\rho \Phi +\Phi^2-\frac{3}{2}\alpha^2\Phi \omega_\rho\omega^\rho-3\Box\Phi=0.\label{eq:traceGravEq}
\end{equation}

The variation of the action (\ref{eq:actionWeyl}) with respect to the scalar field $\phi$ gives rise the following equation of motion of $\Phi$
\begin{equation}
	R-3\alpha\nabla_\rho\omega^\rho -\frac{3}{2}\alpha^2\omega_\rho\omega^\rho +\Phi = 0.\label{eq:scalareq}\
\end{equation}

By multiplying Eq.~(\ref{eq:scalareq}) by $\Phi$, and subtracting it from Eq.~(\ref{eq:traceGravEq}), the generalized Klein-Gordon equation for the scalar field $\Phi$ is obtained as given by
\begin{equation}
\Box\Phi -\alpha\nabla_\rho(\Phi \omega^\rho)=0.
\end{equation}
We also need the equation of motion for Weyl vector field that can be obtained by varying the action (\ref{eq:actionWeyl}) with respect to $\omega_\mu$. The variation will result in
\begin{equation}
4\xi^2 \nabla_\nu F^{\mu\nu}-\alpha^2\Phi \omega^\mu +\alpha \nabla^\mu \Phi =0. \label{eq:WeylvectorEq}
\end{equation}

\subsection{Metric and Field Ansatzs}

In this paper, we investigate the properties of an exact vacuum solution in Weyl geometric gravity.  We consider a static spherically symmetric configuration, with the metric given in a general form by
\begin{equation}\label{eq:spherical}\
	ds^2 = e^{\nu(r)} dt^2 -e^{\lambda(r)}dr^2 - r^2 d\Omega^2,
\end{equation}
where $d\Omega^2 = d\theta^2 +\sin^2\theta d\varphi^2$. The solutions of the field equations of Weyl geometric gravity are dependent on the choice of the Weyl vector field $\omega _\mu$, that can be generally represented as $\omega_\mu =(\omega_0, \omega_1,0, 0)$. Without any loss of the generality, in the following we assumed that $\omega_0=0$. For the case when $\omega_0\neq0$, the numerical study of the field equations was carried out in Ref.~\cite{HT3}.

From the assumption of the form on the Weyl vector it follows that $F_{\mu\nu}\equiv 0$. Then,  by employing this condition in the Weyl vector field equation of motion (\ref{eq:WeylvectorEq}), we obtain
\begin{equation}
\Phi'=\alpha\Phi\omega_1.\label{eq:simplifiedKGEq}\
\end{equation}
Since we can write
\begin{equation}
\Box \Phi = \frac{1}{\sqrt{-g}}\frac{\partial}{\partial x^\mu}\left(\sqrt{-g}g^{\mu\nu}\frac{\partial\Phi}{\partial x^\nu} \right),
\end{equation}
and
\begin{equation}
	\nabla_\mu \omega^\mu = \frac{1}{\sqrt{-g}}\frac{\partial}{\partial x^\mu}\left(\sqrt{-g}\omega^\mu \right),
\end{equation}
respectively, Eq.~(\ref{eq:simplifiedKGEq}) now becomes
\begin{eqnarray}
-\alpha \Phi \frac{1}{\sqrt{-g}}\frac{d}{dr} \left(\sqrt{-g} \omega^1\right)&=&\frac{1}{\sqrt{-g}}\frac{d}{dr}\left(\sqrt{-g}g^{11}\frac{d\Phi}{dr}\right)\nonumber\\
&&-\alpha\omega^1 \frac{d\Phi}{dr}.
\end{eqnarray}

Now we can write the gravitational field equations (\ref{eq:GravEq}) as
\begin{eqnarray}
&& e^{-\lambda}\left(\frac{\lambda'}{r}-\frac{1}{r^2}\right) +\frac{1}{r^2}=\nonumber\\
&&\frac{e^{-\lambda}}{r^2}\bigg(\frac{r^2}{4}e^\lambda \Phi +2\frac{\Phi'r}{\Phi}-\frac{3r^2}{4}\frac{\Phi'^2}{\Phi^2}-\frac{r^2\lambda'}{2}\frac{\Phi'}{\Phi}+r^2\frac{\Phi''}{\Phi} \bigg), \nonumber\\
 \label{eq:GravEq1}\\
 &&e^{-\lambda}\left(\frac{\nu'}{r}+\frac{1}{r^2}\right) -\frac{1}{r^2}=\nonumber\\
&& \frac{e^{-\lambda}}{r^2}\left(-\frac{r^2}{4}e^\lambda \Phi -2\frac{\Phi'r}{\Phi}-\frac{3r^2}{4}\frac{\Phi'^2}{\Phi^2}-\frac{r^2\nu'}{2}\frac{\Phi'}{\Phi} \right),
 \label{eq:GravEq2}\\
 && re^{-\lambda}\left(2\nu'' +v'^2-\lambda'\nu' +\frac{2\nu'}{r}-\frac{2\lambda'}{r} \right)=\nonumber\\
&&-(4-2r\lambda'+2r\nu')\frac{\Phi'}{\Phi}-r\left(e^\lambda\Phi+4\frac{\Phi''}{\Phi}-\frac{\Phi'^2}{\Phi^2}\right).
 \label{eq:GravEq3}
\end{eqnarray}

The left hand side of Eqs. (\ref{eq:GravEq1}), (\ref{eq:GravEq2}) and (\ref{eq:GravEq3}) comes from the Einstein tensor, while the right hand side comes from the contributions of the Weyl geometry. Hence, we can define the effective energy density $\rho$ and pressure $p$ associated  to the scalar field and to the Weyl geometric effects by
\begin{eqnarray}
\rho &=& \frac{e^{-\lambda}}{8\pi r^2}\bigg(\frac{r^2}{4}e^\lambda \Phi +2\frac{\Phi'r}{\Phi}-\frac{3r^2}{4}\frac{\Phi'^2}{\Phi^2}-\frac{r^2\lambda'}{2}\frac{\Phi'}{\Phi} \nonumber\\
&& +r^2\frac{\Phi''}{\Phi} \bigg)=\frac{1}{8\pi}\left[e^{-\lambda}\left(\frac{\lambda'}{r}-\frac{1}{r^2}\right) +\frac{1}{r^2}\right],\\
p &=& -\frac{e^{-\lambda}}{8\pi r^2}\left(\frac{r^2}{4}e^\lambda \Phi +2\frac{\Phi'r}{\Phi}+\frac{3r^2}{4}\frac{\Phi'^2}{\Phi^2}+\frac{r^2\nu'}{2}\frac{\Phi'}{\Phi} \right) \nonumber\\
&=&\frac{1}{8\pi}\left[ e^{-\lambda}\left(\frac{\nu'}{r}+\frac{1}{r^2}\right) -\frac{1}{r^2}\right].\
\end{eqnarray}

By adding Eq.~(\ref{eq:GravEq1}) and (\ref{eq:GravEq2}), we find the evolution equation for the scalar field, given by
\begin{equation}\label{eq:scalarevo}
\frac{\Phi''}{\Phi}-\frac{\lambda'+\nu'}{2}\frac{\Phi'}{\Phi}-\frac{3}{2}\frac{\Phi'^2}{\Phi^2}-\frac{\lambda'+\nu'}{r}=0.
\end{equation}

\section{The Weyl Geometric Black Hole}\label{sect2}

In this Section we will introduce the exact solution of the static spherically symmetric field equations of the Weyl geometric gravity theory, and we will analyze some of its properties.

\subsection{General black hole solutions}\label{subsec:SolType1}

Generally, black hole solutions satisfy the relation $g_{tt}g_{rr}=-1$ between the components of the metric tensor. This case occurs when in a solution to Einstein’s equations the Ricci tensor and the matter energy - momentum tensor have vanishing radial null-null component. On the other hand, the Ricci tensor is proportional to $g_{\mu\nu}$ in the $t-r$ subspace. This also implies that the radial pressure is the negative of the energy density ($p=-\rho$). Hence, in this case, the black hole violates the strong energy condition. The condition $g_{tt}g_{rr}=-1$ also holds if and only if the radial coordinate is an affine parameter on the radial null geodesics \cite{JacobsonCQG2007}.

In a more general approach, one can find black hole solutions in Weyl geometric gravity by assuming $g_{tt}g_{rr}\neq-1$. In our case, we can write that
\begin{equation}
\nu (r) +\lambda (r) = f(r),\label{eq:nulambdaf}
\end{equation}
where $f(r)$ is an arbitrary function of radial coordinate. As a function of the scalar field $\Phi$, one can find the function $f(r)$ from Eq.~(\ref{eq:scalarevo}) as given by
\begin{equation}\label{eq:f(r)integral}
f(r) = \int \frac{2\Phi'' \Phi - 3\Phi'^2}{2\Phi+r\Phi'} rdr.
\end{equation}

This is the general condition that must be satisfied by the scalar field for obtaining black hole solutions in Weyl geometric gravity. Equivalently, we obtain for $\Phi$ the differential equation
\begin{equation}\label{eq:scalarevo}
\frac{\Phi''}{\Phi}-\frac{3}{2}\frac{\Phi'^2}{\Phi^2}-\frac{f'(r)}{2}\frac{\Phi'}{\Phi}-\frac{f'(r)}{r}=0.
\end{equation}

\subsection{Metric of the Weyl geometric black hole}

In this Subsection, we will briefly review the black hole solution found in Ref.~\cite{HT4}, which corresponds to the case when $g_{tt}g_{rr}=-1$. We consider the following condition for the metric tensor potentials
\begin{equation}
\nu(r) +\lambda(r) = 0,~~ \forall r>0.\label{eq:lambdaplusnu}\
\end{equation}

Using this assumption, from Eq.~(\ref{eq:scalarevo}) we immediately obtain the differential equation satisfied by $\Phi$
\begin{equation}
\Phi''=\frac{3\Phi'^2}{2\Phi},
\end{equation}
 corresponding to the choice $f(r)=0$ in Eq.~(\ref{eq:f(r)integral}). The solution of the above equation is
\begin{equation}
\Phi(r)=\frac{C_1}{(r+C_2)^2},\label{eq:scalarsol1}\
\end{equation}
where $C_2$ is just an arbitrary integration constant. The scalar field satisfies the condition $\Phi(r) \rightarrow 0$ at infinity. One can also find the Weyl vector field
\begin{equation}
\omega_1 = \frac{\Phi'}{\alpha \Phi}=-\frac{2}{\alpha(r+C_2)}.\label{eq:Weylvectorsol1}\
\end{equation}
Then, from the gravitational field equation Eq.~(\ref{eq:GravEq1}) one can find the metric potentials  as
\begin{equation}
e^{-\lambda}=e^{\nu}=1-\delta+\frac{\delta(2-\delta)}{3r_g}r -\frac{r_g}{r}+C_3r^2, \label{eq:StarType1}
\end{equation}
where $\delta, r_g, $ and $C_3$ are arbitrary constants and $C_2 = 3r_g/\delta$.

This metric is the generalization of the Schwarzschild-de Sitter solution. There is a Schwarzschild-type singularity at the center due to the presence of the $r_{g}/r$ term in the metric, this term breaks the conformal symmetry spontaneously. Additionally, another conformal breaking term is the integration constant $C_{3}$ which does not exist in the original action. If we choose $C_{3}=0$, the resulting metric will just mimic spacetime in GR but with additional linear term in $r$. For $\delta=0, 2$, the spacetime will become asymptotically flat. The spacetime could contain a BH but immersed in scalar field which generates linear term in the metric.  For negative $C_{3}$, there will be cosmological horizon and the spacetime will be asymptotically de Sitter-like. For positive $C_{3}$, the spacetime could become asymptotically Anti-de Sitter. All of these possibilities of spacetime can contain BH.   

Generically, for $C_{3}<0$, there will be black hole and cosmological horizons. The second term in the solution is the  constant $\delta$ term, which appears to be like a global monopole term~\cite{YuNPB1994}. Similar metrics can also be obtained  for the Kiselev black hole~\cite{KiselevCQG2003}, and in Rastall gravity \cite{SaktiAnnPhys2020,SaktiPhysDark2021}, by assuming some arbitrary values of the black hole's parameters. Furthermore, this metric potential form is also similar to the black hole solution in dRGT gravity \cite{Burikham:2017gdm,Burikham:2020dfi,Ponglertsakul:2018rot,GhoshWongjunEPJC2016}. Nonetheless, we would like to emphasize that $C_3$ comes as an integration constant, related to the scalar field, a situation which is different from the dRGT gravitational theory, where the cosmological constant is present directly from the beginning in the action of the theory. Note that the constants $C_1,C_2,C_3$ and $\delta$ are dependent on each other through the following equation
\begin{equation}
	3C_3C_2^2+\frac{C_1}{4}=3-\delta, \label{eq:C1C2C3rel}
\end{equation}
which is obtained when we solve Eq.~(\ref{eq:GravEq1}) to obtain the metric potentials (\ref{eq:StarType1}) \cite{HT4}.

From Eq.~(\ref{eq:C1C2C3rel}), we find $C_1= 4(3-\delta -27C_3 r_g^2/\delta^2)$. The scalar and vector fields then take the following forms
\begin{eqnarray}
\Phi (r) &=& \frac{4\left(3\delta^2 -\delta^3 -27C_3r_g^2\right)}{(\delta r +3r_g)^2}, \label{eq:scalarwithdelta}\\
\omega_1 (r) &=& -\frac{2\delta}{\alpha(\delta r +3r_g)}.\label{vectorwithdelta}
\end{eqnarray}

The vector field vanishes when $\delta=0$, while the scalar field remains constant. At spatial infinity, both the scalar and the vector fields will vanish.

\subsection{The horizons of the Weyl geometric black hole}

Within this work, we interpret the solution with the metric potentials given by Eq.~(\ref{eq:StarType1}) as representing a hairy black hole solution. Due to the presence of  the constant that mimics a cosmological constant, the Weyl geometric black hole possesses more than two horizons. In the following, we will redefine $C_3 =-1/l^2$. Hence, the horizons are the solutions of the third order algebraic equation
\begin{equation}
\Delta = (1-\delta)r-\frac{\delta(\delta-2)}{3r_g}r^2-r_g-\frac{r^3}{l^2}=0.\label{eq:delta}
\end{equation}

The above equation has real roots (for $1/l^2>0$) if the following conditions are satisfied
\begin{equation}
-\frac{27r_g}{l^2}+2-12\delta+15\delta^2-5\delta^3+2\sqrt{(1-4\delta+2\delta^2)^3}>0,\nonumber\
\end{equation} 
\begin{eqnarray}
0\leq\delta<1-\frac{2\sqrt{7}}{7}. \label{eq:deltaconstraint}\
\end{eqnarray}

\begin{figure}
	\centering
		\includegraphics[width=1.0\linewidth]{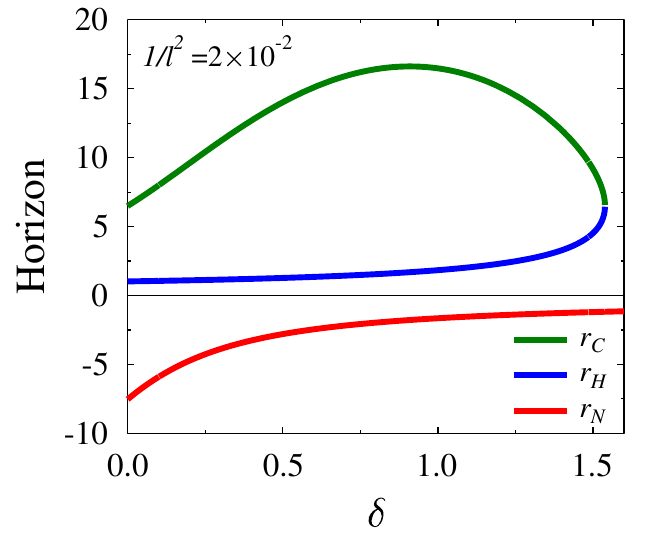}
		\label{fig:horizontype1a}
\caption{The positions of the horizons of the Weyl geometric black hole as a function of $\delta$ for $1/l^2=2\times 10^{-2},r_g=1$. The green, blue, and red lines represent the cosmological, event, and unphysical horizon respectively.}
	\label{fig:horizontype1}
\end{figure}

The plot of the position of horizons versus the parameter $\delta$ is shown in Fig.~\ref{fig:horizontype1}, for $1/l^2=2\times 10^{-2},r_g=1$. The cosmological horizon $r_C$ is given by the green line in Fig.~\ref{fig:horizontype1}. The event horizon $r_H$ is represented by the blue line. There is a third horizon, however, it is unphysical since it is negative horizon ($r_N$) (the red line in Fig.~ \ref{fig:horizontype1}). For cubic equation (\ref{eq:delta}), there are typical analytic expressions for $r_{H}, r_{C}, r_{N}$  which we will not present here. Notably for $\delta > 0$, the cosmological horizon $r_{C}$ can extend far beyond the typical horizon at $r_{C}=l$ in $\delta = 0$ case.

\subsection{Effective density and pressure}

The effective energy density and pressure which appear due to the contribution from the scalar and vector fields in Weyl geometric gravity can be written for the case of the exact black hole solution as
\begin{eqnarray}
\rho &=& \frac{1}{24\pi r^2}\left(3\delta+\frac{2(\delta-2)\delta}{r_g}r+\frac{9r^2}{l^2}\right), \label{eq:rho1}\\
p&=&-\rho. \label{eq:p1}\
\end{eqnarray}

Note that both the density and the pressure are divergent at the center for nonzero $\delta$.  At spatial infinity, the energy density approaches the ``dark energy" density $3/(8\pi l^2)$. Thus, this black hole of Weyl geometric gravity is different from the Schwarzschild-de Sitter solution in GR, since it describes a physical configuration in which there are density and pressure contributions from the Weyl scalar and vector fields.

The pressure vanishes at two different points that can be obtained by solving the equation $p(R)=0$, which gives
\begin{equation}
	R_\pm= \frac{\delta^2-2 \delta  \pm \sqrt{
			-\displaystyle{\frac{27 r_g^2 \delta}{l^2}} + 4 \delta^2 -
			4 \delta^3 + \delta^4}}{-\displaystyle{\frac{9 r_g}{l^2}}}.\label{eq:RfromPressure}
\end{equation}

\begin{figure}
	\centering
	\includegraphics[width=1.0\linewidth]{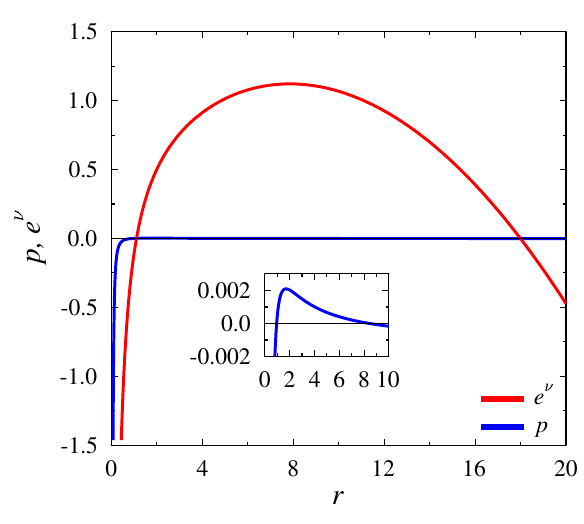}
	\caption{ The metric tensor $e^\nu$ and the effective pressure $p$ of the Weyl geometric black hole. The metric tensor component $e^{\nu}$ is shown as a red line, while the pressure is represented by the blue line. The plots are obtained for $\delta=0.24$, $r_g=1$ and $1/l^2=10^{-2}$.}
	\label{fig:PlotPressMet_vs_r}
\end{figure}

Fig.~\ref{fig:PlotPressMet_vs_r} presents the plot of the effective pressure $p$ and the metric function $e^\nu$ showing the position of the zero-pressure points in comparison to the horizons of the black hole. The plots are obtained for $\delta=0.24, r_g=1$ and $1/l^2=10^{-2}$. With these values of the parameters, we can obtain $p(R)=0 $ at $ R_+=0.948019$ and $R_-=17.9947$. We also have $r_H =1.12597$ and $r_C=8.87784$. Hence, for the given parameters, $R_+< r_H < r_C<  R_-$. The maximum effective pressure occurs at $r=1.70455$, which is outside of $r_H$, but inside of $r_C$ and $R_-$.

\subsection{Dark matter and the Weyl geometric black hole}

In the original paper \cite{HT4}, the effective energy density and pressure associated to the Weyl scalar and vector fields are assumed to represent in an effective form the dark matter that can be found around the baryonic mass distributions in galaxies. Therefore, one can compute the geometric mass that contains the dark matter contribution by integrating the energy density over the volume.

The effective geometric mass of the dark matter for the Weyl geometric black hole solution is given by
\begin{eqnarray}\label{eq:massG}
M_G(r) =\int_0^r 4\pi \rho r'^2 dr' =\frac{r}{2}\left[\delta+\frac{r^{2}}{l^2}-\frac{(2-\delta)\delta}{3r_g}r+\frac{r_g}{r} \right].\nonumber\\
\end{eqnarray}
Note that this mass is not the black hole mass.

We can find the accumulated mass at the points where the pressure vanishes by inserting Eq.~(\ref{eq:RfromPressure}) into the  geometric mass (\ref{eq:massG}) to obtain
\begin{eqnarray}
	M_{G\pm}&=&\frac{r_g}{2}+\frac{\delta\left(3\delta^2-6\delta\pm 2\sqrt{\delta^2(\delta-2)^2+27C_3r_g^2 \delta} \right)}{54C_3r_g}\nonumber\\
	&&+\frac{\delta^2(\delta-2)^2\left(\delta^2-2\delta \pm \sqrt{\delta^2(\delta-2)^2+27C_3r_g^2 \delta}\right)}{729C_3^2r_g^3}.\nonumber\\
	\label{eq:Mmax}
\end{eqnarray}

\begin{figure}
	\centering
	\includegraphics[width=1.0\linewidth]{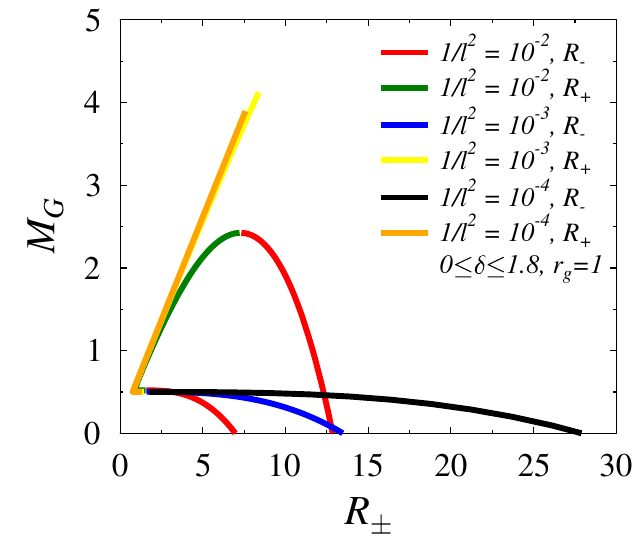}
	\caption{Effective geometric mass $M_G$ is plotted versus $R_\pm$.}
	\label{fig:MR_Relation}
\end{figure}

We plot the effective geometric mass as a function of $R_\pm$ in Fig.~\ref{fig:MR_Relation}. The parameter $r_g$ is given by $r_{g}=1,$ while the geometric mass $M_{G}$ ranges between $0-4$ depending on $\delta, l^{2}$. The geometric mass can reach more than 5 times of the black hole mass for $\delta \gtrsim 1.6$ at the specific radius $R_{+}$. As distance grows, the accumulated mass will continue to increase since the density approaches constant dark energy density $\displaystyle{\frac{3}{8\pi l^{2}}}$ asymptotically at large distance.

It is interesting to consider two cases for the accumulated geometric mass, corresponding to  $R_->R_+$, and $R_- =R_+$, respectively. When $R_->R_+$, we can find the mass difference between $R_{+}$ and $R_{-}$, $\Delta M_G = M_{G-} - M_{G+}$ given by
\begin{equation}
\Delta M_G = -\frac{2l^4\left[\delta^2(\delta-2)^2-\frac{27\delta r_g^2}{l^2}\right]^{3/2}}{729r_g^3}.\label{eq:massGdifference}
\end{equation}
This mass difference acts like the mass of a ``dark matter halo" surrounding the Weyl geometric black hole.

On the other hand, for $R_- =R_+$ we find
\begin{equation}
R=R_-=R_+ = -\frac{\delta(\delta-2)l^2}{9r_g},
\end{equation}
and
\be
M_G= \frac{r_g}{2-\delta},
\ee
respectively.

Note that we have also used the relation $	-27 r_g^2 + (4 \delta -
4 \delta^2 + \delta^3)l^2=0$ from the vanishing square-root term of Eq.~(\ref{eq:RfromPressure}). Since we already have (\ref{eq:deltaconstraint}) for $\delta$, it follows that  $R$ is always positive. In this case, the geometric mass does not depend explicitly on $1/l^2$.

\section{The thermodynamics of the Weyl geometric black hole on the event horizon}\label{sect3}

In the present section we will study the thermodynamics of the Weyl geometric black hole. The first law of black hole thermodynamics thus takes the form
\begin{equation}
	dM_H= T_H dS_{BH} ,\label{eq:thermodynamiclaw1}
\end{equation}
where $M_H$, $T_H$ and $S_{BH}$ are the thermodynamic mass, Hawking temperature, and Bekenstein-Hawking entropy, respectively. Note that even we have a comsological constant-like parameter $C_3=-1/l^2$, we do not need to consider the thermodynamic pressure related to this constant because that constant comes as a consequence of the existence of scalar and Weyl vector fields. So, this constant is different with real cosmological constant that appears in the action likewise in the Einstein general relativity. Such term also appears in the black hole's solution in the dRGT massive gravity theory \cite{Pi}.

\subsection{The Hawking temperature}\label{app:Hawking}

For a generic spherically symmetric spacetime given by (\ref{eq:spherical}), if both $g_{00}=e^{\nu}$ and $1/g_{rr}=-e^{-\lambda}$ have the same zero at $r_{H}$, it can be shown that the Hawking temperature $T_H$ is related to the surface gravity
\begin{equation}
	\kappa_{H} = \frac{g'_{00}}{2}\sqrt{(-g_{00}g_{rr})^{-1}}\Big{|}_{r=r_{H}},
\end{equation}
computed on the event horizon of the black hole, by the relation,
\begin{equation}
	T_H =\frac{\kappa_H}{2\pi}.\label{eq:Hawkingtosurfacegrav}
\end{equation}$\kappa_H$ is the surface gravity .

To show this, we perform the coordinate transformation
\begin{equation}
	g_{rr}dr^{2} \to \frac{dr'^{2}}{g_{00}(r(r'))}.
\end{equation}
Then the Hawking temperature is given by the conventional formula
\begin{equation}\label{TH1}
	T_H = \frac{g'_{00}(r')}{4\pi} = \frac{g'_{00}(r)\sqrt{(-g_{00}g_{rr})^{-1}}}{4\pi},
\end{equation}
where, in the new coordinates,
\be
\kappa_{H}=\frac{1}{2}g'_{00}(r')=\frac{1}{2}g'_{00}(r)\sqrt{(-g_{00}g_{rr})^{-1}},
\ee
at $r_{H}$. This ends the proof of this result. Here, the surface gravity is defined as the acceleration experienced by an object at the event horizon of the black hole.

Alternatively, in terms of the Killing vector, the surface gravity can be given by
\begin{equation}
	\kappa = \sqrt{-\frac{1}{2}\nabla^\nu \xi^\mu \nabla_\nu \xi_\mu}.\label{eq:surfacegravitygeneral}\
\end{equation}

Here $\xi^\mu$ denotes the Killing vector possessed by the spacetime metric.  For the spherically symmetric spacetime (\ref{eq:spherical}), we have $\xi^\mu =(1,0,0,0)$. Hence, the non-vanishing components of the covariant derivatives of the Killing vector are
\begin{eqnarray}
	\nabla^t \xi^r \nabla_t \xi_r = \nabla^r \xi^t \nabla_r \xi_t=-\frac{{\nu'} ^2}{4}e^{\nu -\lambda}.   \label{Keqn}
\end{eqnarray}
Thus, we obtain the surface gravity and the Hawking temperature on the event horizon as
\begin{equation}
	\kappa_H =\frac{\nu'}{2}e^{(\nu-\lambda)/2}\bigg|_{r=r_H}, ~~~ T_H = \frac{\nu'}{4\pi}e^{(\nu-\lambda)/2}\bigg|_{r=r_H}.\label{eq:surfacegravityforspherical}\
\end{equation}

This is the same result as presented in (\ref{TH1}). Under the condition that both $e^{\nu}$ and $e^{-\lambda}$ contain the same zeros at $r=r_{H}$, and by using the l'Hopital rule, we find the relation $\nu'|_{r=r_{H}} = -\lambda'|_{r=r_{H}}$. Hence, one can also rewrite the equation for the Hawking temperature as
\begin{equation}
	T_H = \frac{1}{4\pi}\left(-\nu'\lambda'e^{\nu-\lambda}\right)^{1/2}\bigg|_{r=r_H}. \label{THeqn}
\end{equation}

This is exactly the equation for the Hawking temperature obtained from the tunneling method \cite{SiahaanEPJC2016}. Note that this formula is valid generically {\it regardless} of the vacuum condition $\nu + \lambda = 0$, as long as {\it both} metric tensor components $g_{00}$ and $g_{rr}$ have the same zeros at the horizon.

\begin{figure}
	\includegraphics[width=1.0\linewidth]{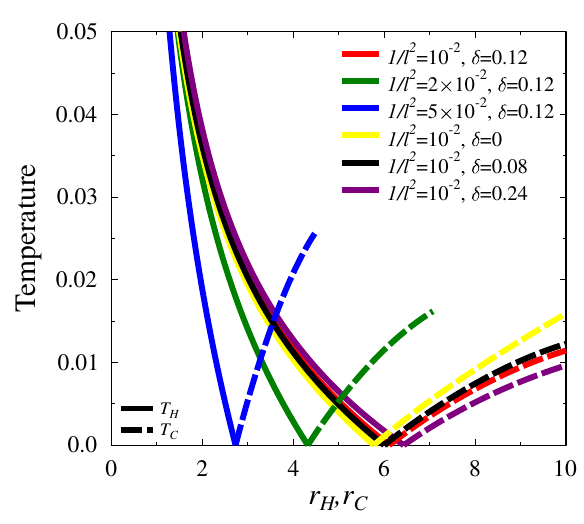}
	\caption{Hawking temperature for various values of the parameters of the Weyl geometric black hole. The solid lines represent Hawking temperature of $r_H$ while the dashed lines represent Hawking temperature of $r_C$. At $r_H=r_C=r_{Nar}$, the temperature vanishes.}
	\label{fig:PlotHawking}
\end{figure}

For the Weyl geometric black hole, the Hawking temperature is
\begin{equation}
	T_H = \frac{1}{4\pi}\left[\frac{r_g}{r^2_H}-2\frac{r_H}{l^2} -\frac{\delta(\delta-2)}{3r_g}\right] .\label{eq:Hawkingtemp_rH}\
\end{equation}
Note that this Hawking temperature of the Weyl geometric black holes is not identical to the temperature of a  black hole in dRGT gravity~\cite{GhoshWongjunEPJC2016}. The plot of the Hawking temperature is presented in Fig. \ref{fig:PlotHawking}. The plot is divided into two region which are $0\leq r_H \leq r_{Nar}$ and  $r_{Nar}\leq r_C \leq l$ where $r_0$ is the Nariai limit where $r_H=r_C=r_{Nar}$. Near zero, the Hawking temperature is very large. In Nariai limit, the Hawking temperature will vanish. Then there will exist the temperature in the cosmological horizon. The thermodynamics on the cosmological horizon will be given in the next section.

We will also see that the entropy of the Weyl geometric black holes is not similar to that of the black holes in the dRGT gravity, due to the presence of the non-minimal coupling between gravity and the scalar field.

For $\delta=0$, Eq.~(\ref{eq:Hawkingtemp_rH}) reduces to
\begin{equation}
	T_H = \frac{1}{4\pi}\left(\frac{r_g}{r^2_H}-\frac{2r_H}{l^2} \right) .\label{eq:HawkimgtempSchdS}\
\end{equation}
It can be seen easyly that above equation gives exactly the Hawking temperature of the Schwarzschild-de Sitter black hole.

\subsection{Entropy from the Noether charge}\label{app:entropy}

The Noether charge formalism provides a systematic and powerful way to calculate conserved quantities, e.g., the entropy, without being tied to the specific details of the gravitational theory. In particular, this formalism is useful to understand black holes, and their thermodynamic properties. The main ingredient of the Noether charge formalism is the diffeomorphism invariance.

For general Weyl invariant theories, the Noether charge formalism was introduced in in Ref.~\cite{Alonso-SeranoPRD2022}. Comparing to that theory, in Weyl geometric gravity, there is no cosmological constant appears in the Lagrangian so that we do not need to include cosmological constant term in the conserved charge.

Interestingly, the Weyl geometric gravity theory, described by the Lagrangian (\ref{eq:actionWeyl}), contains the Ricci scalar coupled to a scalar field, similarly to  the case of the Brans-Dicke theory. We may interpret the other terms in the Lagrangian as the matter contribution. In the Brans-Dicke theory, the entropy does not satisfy the exact area law, since the scalar field and gravity are coupled non-minimally~\cite{VollickPRD2007}. The entropy expression has additional scalar field terms. One may remove these factors by working in the Einstein frame. However, this is not the case that we will investigate within this paper.

We can choose $\xi^{2}$ in a particular way so that we can write the Lagrangian in (\ref{eq:actionWeyl}) in the simple form
\begin{equation}
	\mathcal{L} = \frac{\Phi}{16\pi} R + \mathcal{L}_M, \label{eq:Lagrangian}
\end{equation}
where $\mathcal{L}_M$ represents all the other terms in Eq.~(\ref{eq:actionWeyl}), and which can be interpreted  as the Lagrangian of an effective form of matter. The matter terms do not contribute to the entropy directly through the Noether charge. Hence, the non-vanishing contribution to the entropy in the Noether charge formalism comes only through the coupling of the gravitational field with the scalar, and which is given by~\cite{VollickPRD2007}
\begin{equation}
	Q=\int _{\partial\Sigma}Q^{\mu\nu}d\sigma_{\,u\nu}, \label{eq:Noether}
\end{equation}
where
\begin{eqnarray}
	Q^{\mu\nu}&=&\frac{\Phi}{16\pi}(\nabla^\mu \xi^\nu -\nabla^\nu \xi^\mu), \\
	d\sigma_{\mu\nu} &=& \frac{1}{4}\sqrt{-g}\varepsilon_{\mu\nu\theta\phi}\text{det}\left(\frac{\partial(x^\mu,x^\nu)}{\partial(\theta,\phi)}\right)d\theta d\phi.\label{eq:dsigma}
\end{eqnarray}
$d\sigma_{\mu\nu}$ is the surface element of a closed two-dimensional surface of the event horizon, and $\partial\Sigma$ is a two-sphere of an arbitrary radius, which  we choose to be the horizon.

We can now compute the conserved charge $Q$ for the spacetime metric (\ref{eq:spherical}). The non-vanishing components of the covariant derivative of the Killing vector are already given in Eq.~(\ref{Keqn}). The element of the two-dimensional surface is given by
\begin{equation}
	d\sigma_{tr} = e^{(\nu+\lambda)/2}r^2\sin^2\theta d\theta\wedge d\phi|_{r=r_{H}}.
\end{equation}
Using these expressions into the integral form of $Q$, we obtain
\begin{equation}
	Q|_{r=r_H}=\frac{\Phi_H e^{\frac{\nu-\lambda}{2}}\nu'}{4} r_H^2 .\label{eq:Qgeneral}\
\end{equation}

Then the entropy of the Weyl geometric black hole is given by
\begin{equation}\label{eq:entropyfromNoether}
	S_{BH}= \frac{2\pi Q|_{r=r_H}}{\kappa_H} = \Phi_H\pi r_H^2.
\end{equation}

\begin{figure}
	\centering
	\includegraphics[width=1.0\linewidth]{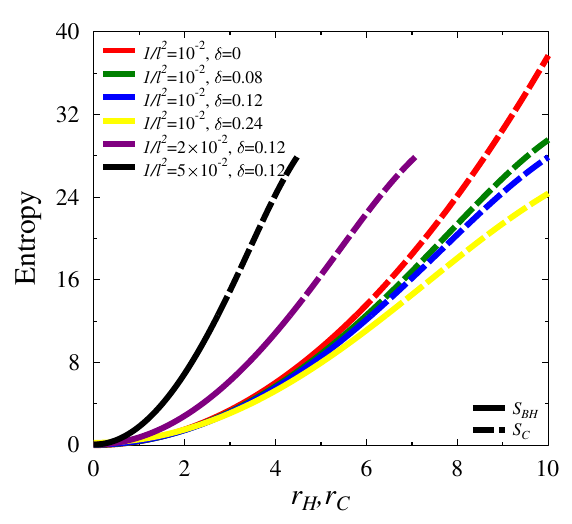}
	\caption{Weyl geometric black hole's entropy for various values of the parameters of the Weyl geometric black hole. The solid lines represent Bekenstein-Hawking entropy on $r_H$ while the dashed lines represent the entropy on $r_C$.}
	\label{fig:PlotEntropy}
\end{figure}

Note that $\Phi_H$ is the value of the scalar field on the event horizon. Yet, the entropy (\ref{eq:entropyfromNoether}) is the general entropy for the Lagrangian (\ref{eq:Lagrangian}). It differs from the area law for the entropy by a scalar factor $\Phi_{H}$ at the horizon. The spacetime metric of the Weyl geometric black hole is identical to the black hole solution in dRGT theory. However, we obtain a different entropy from the dRGT theory, which was computed in Ref.~\cite{GhoshWongjunEPJC2016}, because the Lagrangian is totally different from the Lagrangian of the dRGT gravity. The plot of the black hole entropy is shown in Fig. \ref{fig:PlotEntropy}. Unlike the Hawking temperature, in Nariai limit, the entropy does not vanish. The entropy continues to exist from $r_H=0$ to the de Sitter length.

It would be also interesting to study the entropy of the Weyl geometric black hole in the Einstein frame, in which the scalar field is not coupled directly to the Ricci scalar. We leave this issue for future study.

For the special case when $\delta=0$, the scalar field on the event horizon is
\begin{equation}
	\Phi_H =\frac{12}{l^2}.\label{eq:scalardelta0}\
\end{equation}
The entropy is then
\begin{equation}
	S_{BH}=\frac{12\pi r_H^2}{l^2}.\label{eq:entropydelta0}\
\end{equation}

The entropy (\ref{eq:entropydelta0}) is proportional to the black hole horizon area, yet with an extra factor rather than $1/4$. Hence, we can see that the entropy will vanish for $1/l^2 =0$. This entropy does not reproduce the entropy of the Schwarzschild solution for a vanishing cosmological constant-like term. The entropy expression (\ref{eq:entropydelta0}) is obviously different from the one obtained from the Schwarzschild-de Sitter solution in GR.

In Weyl geometric gravity, in order to obtain the GR limit, one needs to break the conformal symmetry, e.g., by introducing a  spontaneous symmetry breaking, which can be achieved  by giving $\Phi$ a vacuum expectation value $\Phi_{0}(r)$~\cite{Ghilencea:2021lpa}, which satisfies (\ref{eq:scalarsol1}) for the spherically symmetric vacuum. In the conformal-breaking vacuum, the gravitational constant will vary with $r$ as $G = 1/\Phi_{0}(r)$.

In the GR limit, the entropy of the Weyl geometric black hole for $\delta \to 0$ will reduce to the Schwarzschild-de Sitter black hole entropy $S=\pi r_{H}^{2}/G$, with the identification $G=1/\Phi_{H}$.

\subsection{The black hole masses}

The thermodynamic mass of the black hole is different from the geometric mass which we have introduced in  Eq.~(\ref{eq:massG}). Beside the effective geometric mass of the dark matter, in the following, we will consider two different masses which are the mass function related to the integration constant $r_g$ and the thermodynamic mass.  In the Weyl geometric black hole solution, we have the gravitational radius $r_g$ of the black hole, which is proportional to the ``gravitational" mass, i.e., $r_g=2GM$ where $M$ is the black hole ``gravitational" mass and $G\equiv 1/\Phi_{H}$. This choice is similar to the way the mass of the black hole is obtained in General Relativity and Brans-Dicke theory \cite{XuLuYangJiangEntropy2023}.

For asymptotically de Sitter solutions, one way to compute the black hole~(gravitational) mass has been described in Ref.~\cite{DolanCQG2019}. In this approach, the invariant definition of the mass for a black hole in de Sitter spacetime does not rely on taking the limit $r\rightarrow\infty$. All what is necessary for the the calculation of the mass is the existence of a time-like Killing vector at some point $r < r_C$,  in a region of spacetime containing a 2-sphere surrounding the mass. The mass of the black hole can then be calculated by integrating over a sphere of any radius, as long as it thoroughly surrounds the mass, and is within a region containing the time-like Killing vector. Similar to the Schwarzschild-de Sitter solution \cite{DolanCQG2019}, one can then find the gravitational mass of Weyl geometric black hole as $M=\Phi_{H}r_g/2$.

In terms of the event horizon, we can solve Eq.~(\ref{eq:delta}) to find the black hole gravitational mass as a function of $r_{H}$,
\begin{eqnarray}
\hspace{-0.5cm}M_\pm&=& \Phi_{H}\frac{1}{12}\Bigg[3\left(-\frac{r_H^3}{l^2}-\delta r_H+r_H\right) \nonumber\\
\hspace{-0.5cm}&\pm& \left. \sqrt{9\left(-\frac{r_H^3}{l^2}-\delta r_H+r_H\right)^2-12r_H^2\delta(\delta-2)}\right].\label{eq:massinrH}\
\end{eqnarray}

Note that in terms of the event horizon the mass has two branches. Thus, the mass could be negative when one adopts the solution $M_-$, if the value in the square root is bigger than the other terms. Hence, for this case, we need the condition $\delta>2$ in order for the mass to be positive. However, $\delta$ has a constraint given by Eq.~(\ref{eq:deltaconstraint}), where the maximum value of $\delta$ is $0.244071$. Therefore the negative sign of the mass does not obey this condition, and we will not consider this case further.

In addition to the black hole mass, the geometric mass on the event horizon can also be computed. Interestingly enough, by using Eq.~(\ref{eq:massG}) and (\ref{eq:delta}) at $r=r_{H}$, a very simple formula for the geometric mass on the event horizon can be obtained for the Weyl geometric black hole
\begin{equation}
M_{G}|_{r=r_H}= \frac{r_H}{2}.\label{eq:massGonrH}\
\end{equation}
This condition is identical to the Schwarzschild radius $r_H=2M_G$ for the Schwarzschild black hole.

\begin{figure}
	\centering
	\includegraphics[width=1.0\linewidth]{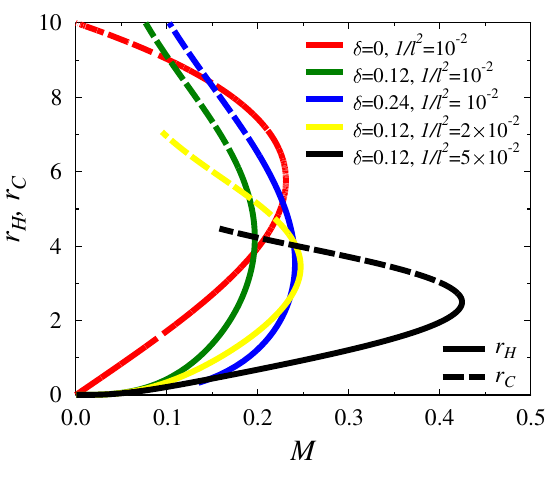}
	\caption{The Weyl geometric black hole mass versus $r_{H}$ and $r_{C}$. }
	\label{Mrhc}
\end{figure}

There is a maximum mass $M$ that can be obtained from $dM/dr_H=0$. Due to the factor $\Phi_{H}$ which intrinsically depends on $M$, the maximum gravtiational mass does not occur at the Nariai limit as shown in Fig.~\ref{Mrhc}, it occurs at a bit smaller radius. On the other hand if we assume $G_{\rm eff}=1/\Phi_{H}$ to be {\it constant}, the maximum mass will be given by
\begin{equation}
M_{\rm max}=\Phi_{H}\frac{l\sqrt{(\delta +1)^3}+(1-2 \delta ) l\sqrt{\delta +1} }{6 \sqrt{3}}.\label{eq:maximumM_rH}
\end{equation}
The position of the horizon for the maximum mass is obtained as
\begin{equation}
r_{Nar}= \sqrt{\frac{(\delta+1)l^2}{3}},\label{eq:positionMmax}
\end{equation}
which is the Nariai limit where where event horizon and cosmic horizon merge. In this case, the maximum mass is simply the mass of black hole in the Nariai limit. This is shown in Fig.~\ref{Mrhc}.

The second mass we consider in the following is the thermodynamic mass which satisfies the first law of thermodynamics. From Eq. (\ref{eq:thermodynamiclaw1}), we can find
\begin{equation}
M_H =\int T_H dS_{BH}. \label{eq:integralTdS}
\end{equation}
To compute above integral, we use the mass function (\ref{eq:massinrH}) in the $T_H$ and $S_{BH}$ and the relation $dS_{BH}=(\partial S_{BH}/\partial r_H)dr_H$. Finally, we can obtain
\begin{eqnarray}
M_H &=& \frac{r_H}{l^2}\sqrt{3\left(\frac{3r_H^4}{l^4}+\frac{6r_H^2}{l^2}(\delta-1)+\delta(2-\delta)+3\right)} \nonumber\\
&-&\frac{3r_H^3}{l^4}+\frac{3r_H}{l^2}-\frac{3r_H}{l^2}\delta.\label{eq:newmassinrH}
\end{eqnarray}
The position of the maximum thermodynamic mass is at the Nariai limit given by Eq. (\ref{eq:positionMmax}).  The maximum thermodynamic mass is given by
\begin{equation}
M_{H max}=\frac{2\left[\sqrt{(\delta +1)^3}+(1-2 \delta ) \sqrt{\delta +1}\right] }{ l\sqrt{3}}.\label{eq:maximumMH_rH}
\end{equation}

\begin{figure}
	\centering
	\includegraphics[width=1.\linewidth]{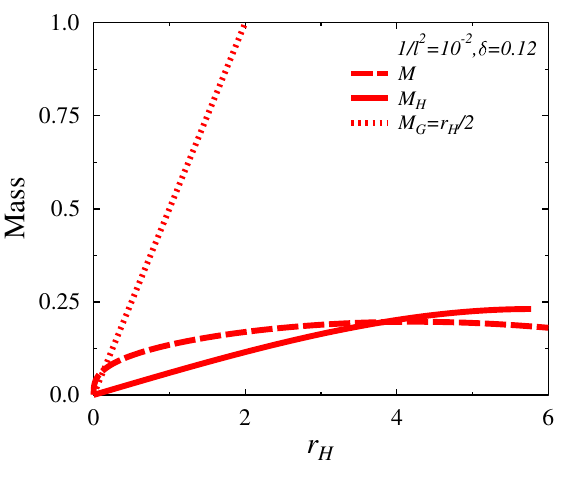}
	\caption{ The masses of Weyl geometric black hole plotted as a function of $r_H$, for different values of the parameters. }
	\label{fig:PlotMassvsrH}
\end{figure}

The black hole gravitational mass $M$, thermodynamic mass $M_H$, and geometric mass $M_G$ are plotted as a function of $r_H$ for the variation of $\delta$ and $1/l^2$ in Fig.~\ref{fig:PlotMassvsrH}. 
We can see from Fig. \ref{fig:PlotMassvsrH} that the position of the maximum mass for $M_H$ are at the Nariai limit while the maximum of $M$ is at smaller radius.  However, the geometric mass at $r_{H}$ is quite different since it is only the linear function of $r_H$ and does not depend on $\delta,l^2$. Variation of $\delta$ gives opposite effect to both masses. Remarkably, for $\delta = 0,$ the gravitational mass $M$ and thermodynamic mass $M_{H}$ become identical. 
\begin{equation}
M_H|_{\delta=0} = M|_{\delta=0} = \frac{6}{l^2}\left(r_H-\frac{r_H^3}{l^2} \right). \label{eq:thermomassinrHnodelta}
\end{equation}

\section{Thermodynamics on the cosmological horizon}\label{sect5}

Generically, Weyl geometric black holes could have a cosmological horizon and we can compute the thermodynamic quantities on this horizon. The thermodynamic quantities on the cosmological horizon are obtained by considering the event horizon as the boundary \cite{GomberoffPRD2003}. On the other hand, the thermodynamic quantities on the event horizon are obtained when the cosmological horizon is considered as the boundary.

The plot of the Weyl geometric black hole mass $M$ versus $r_{H}, r_{C}$, is shown in Fig.~\ref{Mrhc}. For each $M$ there could possibly be two horizons up until the maximum mass where $r_{H}=r_{C}$ at the Nariai limit. This limit corresponds to the largest possible Weyl geometric black hole.

The first law of the thermodynamics on the cosmological horizon takes a similar form with that on the event horizon. It is given by
\begin{equation}
	dM_C= T_C dS_{C}.\label{eq:thermodynamiclaw2}\
\end{equation}

First, the Hawking temperature computed from Eq.~(\ref{THeqn}), and the entropy, obtained from Eq.~(\ref{eq:entropyfromNoether}), are given by
\begin{eqnarray}
	T_C&=& -\frac{1}{4\pi}\left(\frac{r_{g}}{r^2_C}-2\frac{r_C}{l^2} -\frac{\delta(\delta-2)}{3r_{g}}\right), \label{eq:Hawkingtemp_rC}\\
	S_{C}&=& \Phi_C\pi r_C^2,\label{eq:entropyfromNoether_rC}\
\end{eqnarray}
where $\Phi_C =\Phi(r_C)$. The temperature of the cosmological horizon is negative since the entropy at the cosmological horizon increases when the black hole mass decreases~\cite{Klemm:2004mb,Cvetic:2018dqf}. 
\begin{figure}
	\centering
	\includegraphics[width=1.\linewidth]{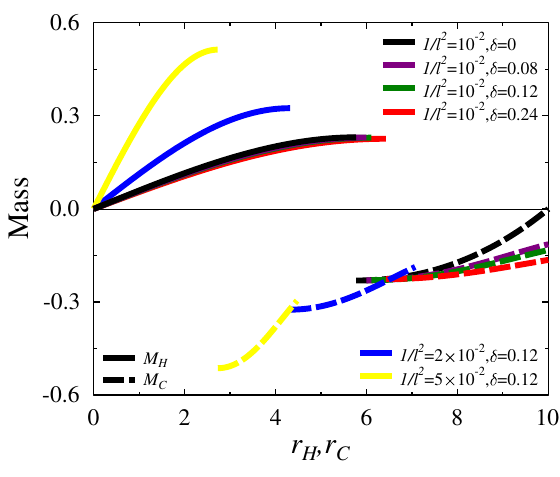}
	\caption{The thermodynamic mass of Weyl geometric black hole for various values of the parameters. The solid lines represent the thermodynamic mass on $r_H$ while the dashed lines represent the thermodynamic mass on $r_C$.}
	\label{fig:PlotMassTot}
\end{figure}

The mass on the cosmological horizon is obtained as
\begin{eqnarray}\label{eq:massinrC}
	M_C&=&\int T_C dS_C \nonumber\\
	&=&-\frac{r_C}{l^2}\sqrt{3\left(\frac{3r_C^4}{l^4}+\frac{6r_C^2}{l^2}(\delta-1)+\delta(2-\delta)+3\right)} \nonumber\\
	&+&\frac{3r_C^3}{l^4}-\frac{3r_C}{l^2}+\frac{3r_C}{l^2}\delta.\label{eq:newmassinrC}
\end{eqnarray}
We can see the main difference from the mass on the event horizon, i.e., the overall sign of $M_{C}$ becomes negative~\cite{GomberoffPRD2003}. The plot of the mass from $r_H=0$ to de Sitter length~$L$) is presented in Fig. \ref{fig:PlotMassTot}. It is clear that $M_{C}$ is negative. We expect that when the temperature approaches minimum value and the thermodynamic mass approaches maximum value on the event horizon, the small black hole increases to the Nariai condition at $r_H =r_{Nar}$ by absorbing radiation. On the other hand, the cosmological horizon decreases to the minimum value by emitting radiation. This is the Hawking-Page transition to achieve a large and stable black hole in de Sitter space.

For a system containing two different horizons, the event and cosmological horizons, the total entropy is the sum of the entropies from each horizon~\cite{UranoSaidaCQG2006}
\begin{equation}
	S_{tot} = S_{BH} + S_C. \label{eq:totalentropy}\
\end{equation}

\section{Thermodynamic stability}\label{sect4}

One can study the thermodynamic global stability of the black holes by using the Helmholtz free energy.  On the other hand, to study the local stability of the thermal equilibrium between the black hole, and its surroundings, we can consider the heat capacity of the black hole. The heat capacity is also an important avenue for investigating critical phenomena, and the phase structures of the black hole~\cite{DaviesProcRSocA1977}. More specifically, when the heat capacity diverges or changes sign, it indicates a phase transition, and the breakdown of the equilibrium thermodynamic description. It is worth noting that the local stability does not imply the global one.  

\subsection{The free energy}

Since in this black hole solution we do not consider any particle transfer, we will consider the Helmholtz free energy to study the thermodynamic stability. The Helmholtz free energy is given by
\begin{eqnarray}
F &=& M_H- T_H S_{BH}, \label{eq:Helmholtz}\
\end{eqnarray}
where we use the thermodynamic mass in this investigation.

As compared to the vacuum case, the global thermodynamic stability of a black hole is determined by the condition $F \leq 0$. In this case, $F=0$ is the condition for the first-order phase transition from the vacuum state to the black hole. For the Weyl geometric black hole, in terms of $r_H$, the Helmholtz free energy  is given by
\begin{eqnarray}
F &=& \frac{3r_H^3}{2l^4}+\frac{2r_H(\delta+3)}{2l^2}-\frac{3r_H(\delta+1)(\delta-3)}{6r_H^2} \nonumber\\
&-&\frac{1}{2l^2} \sqrt{\frac{9r_H^6}{l^4}+\frac{18r_H^4(\delta-1)}{l^2}-3r_H^2(\delta+1)(\delta-3)}\nonumber\\
&-&\frac{\delta+3}{6r_H^2} \sqrt{\frac{9r_H^6}{l^4}+\frac{18r_H^4(\delta-1)}{l^2}-3r_H^2(\delta+1)(\delta-3)}.\nonumber\\
& &\
\end{eqnarray}

\begin{figure}
	\centering
	\includegraphics[width=1.0\linewidth]{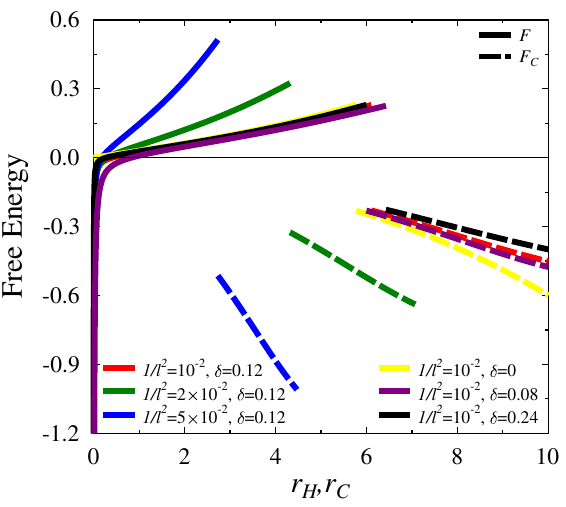}
	\caption{The free energy~$F$ versus $r_H$ and $r_{C}$ for Weyl geometric black hole. }
	\label{fig:PlotHelmholtz}
\end{figure}

Note that the vacuum state without a black hole in this background is obtained when $\delta=r_{H}=0$, which naturally implies $F=0$ as the Schwarzschild-dS black hole. Fig.~\ref{fig:PlotHelmholtz} presents the plots of the Helmholtz free energy versus $r_{H},r_C$. From the range of parameters we choose, it turns out that the Weyl geometric black hole has negative free energy associated with the event horizon for $r_{H}\ll 1$ but positive free energy for large black hole.  For the pure de Sitter space, it is obvious that $F_C<0$ denoting that it is in stable configuration.

\begin{figure}
	\centering
	\includegraphics[width=1.0\linewidth]{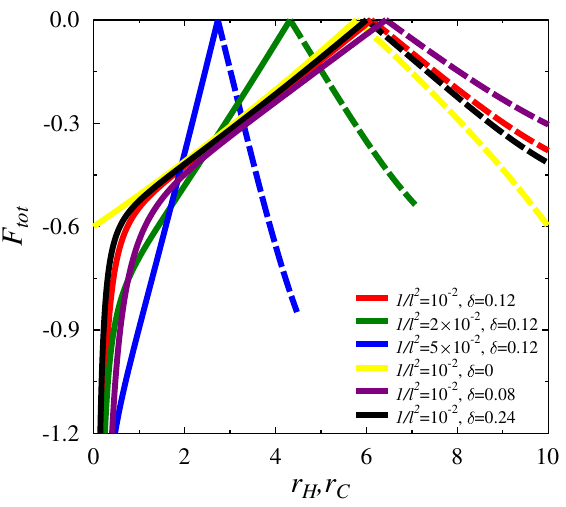}
	\caption{The free energy~$F$ versus $r_H$ and $r_{C}$ for Weyl geometric black hole. }
	\label{fig:PlotHelmholtzTot}
\end{figure}

The thermodynamic stability of a spacetime with two horizons occurs when the total free energy,
\begin{equation}
F_{tot}=F+F_{C}
\end{equation}
where $F_{C}=M_{C}-T_{C}s_{C}$, is negative as shown in Fig.~\ref{fig:PlotHelmholtzTot}. In the Nariai limit $r_{H}=r_{C}$, since $M= -M_{C}, T_{H}= -T_{C}, S_{H} = S_{C}$, the total free energies vanish.

\subsection{Heat capacity}

Now we will study the thermodynamic stability locally by using the heat capacity. The heat capacity on the event horizon can be computed from the relation
\begin{equation}
C_H = \frac{dM_H}{dT_H}=\frac{dM_H}{dr_{H}}\Big/\frac{dT_{H}}{dr_{H}}. \label{eq:heatcap}\
\end{equation}

\begin{figure}
	\centering
	\includegraphics[width=1.0\linewidth]{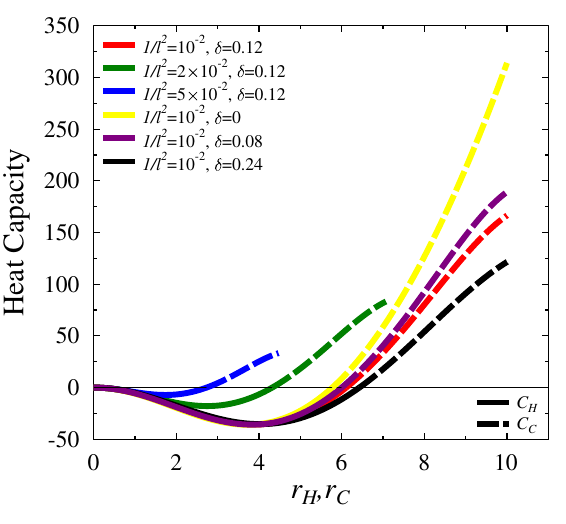}
	\caption{The heat capacity of Weyl geometric black holes versus $r_H$ and $r_{C}$. }
	\label{fig:PlotHeatCapacity}
\end{figure}

The plot of the heat capacity is given in Fig.~\ref{fig:PlotHeatCapacity}. When $C_H$ is positive, the thermal equilibrium with the surrounding environment is stable, while the negative values implies locally thermodynamical instability. The Weyl geometric black hole is locally unstable for small $r_H$, and there is no divergence. The local instability for small black holes only implies that these black holes will evaporate by means of the Hawking radiation and gets hotter. There is a minimum of heat capacity for the Weyl black hole. For the de Sitter space, the heat capacity shows the local stability where there is a maximum of heat capacity at $r_C=l$. On the Nariai limit, both heat capacities are connected smoothly without divergence.

For a continuous $F$ as well as for its first derivatives, the divergence or discontinuity of the heat capacity implies a second-order phase transition, since heat capacity is related to the second derivative of the free energy~\cite{DaviesProcRSocA1977}. For the Weyl geometric black hole, there are no divergences in the heat capacity, and thus no second-order phase transition can take place.

\subsection{Luminosity and evaporation time of Weyl geometric black holes}\label{sec:evap}

The luminosity of the black hole can be approximated using the Stephan-Boltzmann law, or by explicit calculation of evaporation rates \cite{BambiEPJC2018}. It is given by
\begin{equation}
L = \frac{dM_H}{dt}= \sigma A_H T_H^4,\label{eq:luminosity}
\end{equation}
where $\sigma$ is a Stefan–Boltzmann-like constant that depends on the black hole mass and on the particle content of the theory. $A_H$ is the area of the black hole on the horizon, which is given by
\begin{equation}
A_H = \int^{2\pi}_0 \int^{\pi}_0 \left(g_{\theta\theta}g_{\phi\phi}\right)^{1/2} d\theta d\phi.
\end{equation}

For the Weyl geometric black hole,  $A_H = 4\pi r_H^2$. Hence, the black hole luminosity can be written as
\begin{equation}
L= \frac{\sigma r_H^2}{64\pi^3}\left(\frac{r_{g}}{r_H^2}-\frac{2r_H}{l^2}-\frac{\delta(\delta-2)}{4r_{g}}\right)^4.\label{eq:luminosity1}
\end{equation}

\begin{figure}
	\centering
	\includegraphics[width=1.0\linewidth]{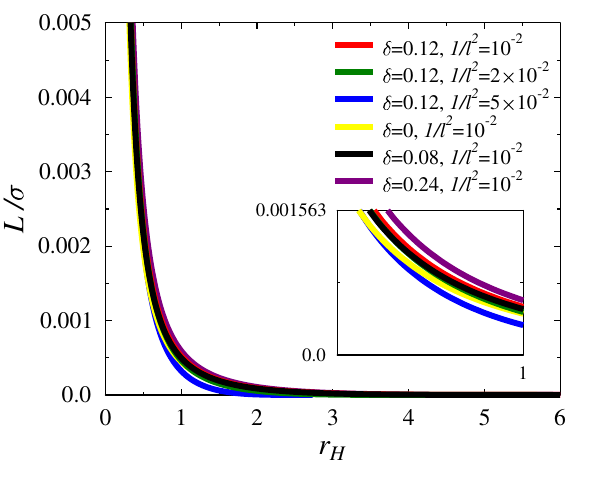}
	\caption{Luminosity of the Weyl geometric black holes versus the radius of the event horizon $r_{H}$.}
	\label{fig:luminosity}
\end{figure}

We show the plot of the luminosity divided by $\sigma$ in Fig. \ref{fig:luminosity}, where we have used the mass function (\ref{eq:massinrH}). The effects of the parameter $1/l^2$ and $\delta$ do not really affect the luminosity.

\subsubsection{The evaporation time}

From Eq.~ (\ref{eq:luminosity}), we can also compute the evaporation time of the black hole. It is given by
\begin{equation}
t_{\rm Evap} = \int_0^{M_H} \frac{d\hat{M}_H}{\sigma A_H T_H^4}.\label{eq:tevap}
\end{equation}

\begin{figure}
	\centering
	\includegraphics[width=1.0\linewidth]{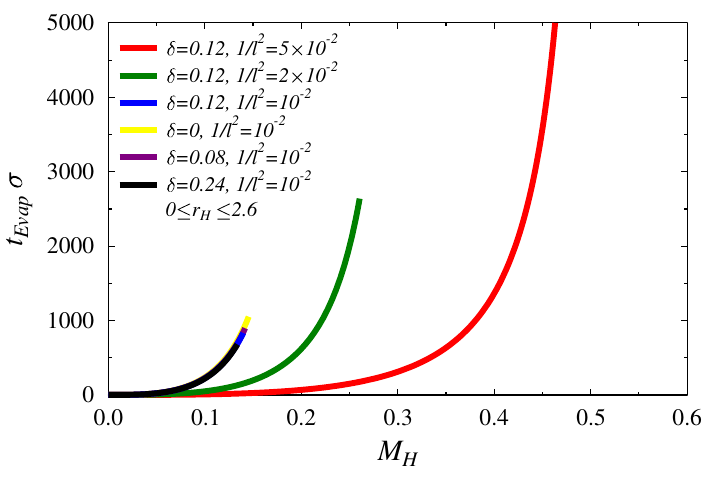}
	\caption{Evaporation time $t_{\rm Evap}$ of the Weyl geometric black hole versus thermodynamic mass $M_H$.}
	\label{fig:tEvap}
\end{figure}
\begin{figure}
	\centering
	\includegraphics[width=1.0\linewidth]{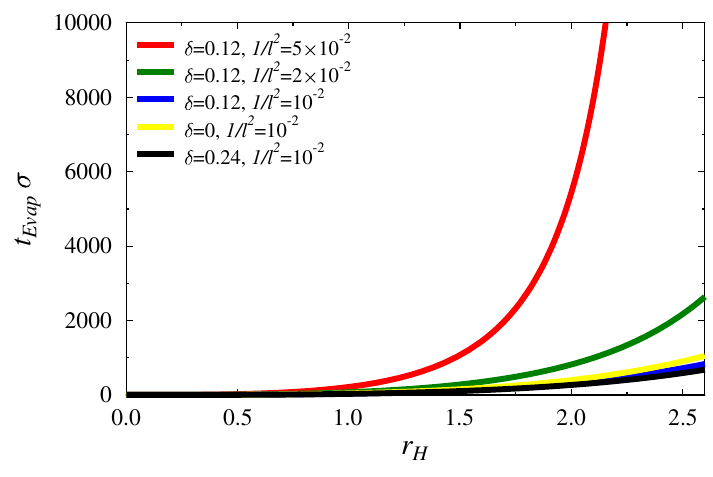}
	\caption{Evaporation time $t_{\rm Evap}$ of the Weyl geometric black hole versus event horizon $r_H$.}
	\label{fig:tEvaprH}
\end{figure}

We show the plot of the evaporation time $t_{\rm Evap}$ of Weyl geometric black hole versus the mass $M_H$ for a certain range of $r_H$ in Fig.~\ref{fig:tEvap}. The evaporation time of Weyl geometric black hole for the given parameter increases with the thermodynamic mass. In the Nariai limit, where $r_{H}\to r_{C}$, the Hawking temperature becomes zero, and $t_{\rm Evap}\to \infty$. Also as shown in Fig. \ref{fig:tEvaprH}, when $r_H$ becomes larger, the evaporation time becomes very large.

\section{Discussions and final remarks}\label{sect6}

Weyl geometric gravity is an interesting extension of general relativity, which implements the requirement of conformal invariance at both the action and field equations level. The starting point is the action constructed in Weyl geometry which contains only two terms. An important result of the theory is the extraction of a scalar degree of freedom through the linearization of the action in terms of the Weyl (Ricci) scalar, and the formulation of then theory as an equivalent scalar-vector-tensor theory. The theory admits an exact vacuum solution in spherical symmetry, corresponding to a particular choice of the Weyl vector, assumed to have only a radial component. As a result the strength of the Weyl vector $\tilde{F}_{\mu \nu}$ identically vanishes. Moreover, the Weyl vector becomes the gradient of the scalar field, $\omega _1=(1/\alpha)d \ln \Phi/dr$. Thus, the solution is not given in the proper Weyl geometry, but in a subclass of it, called integrable Weyl geometry \cite{WIG}. However, it is important to point out that Weyl Integrable Geometry does not reduce to Riemann geometry, since the group of geometrical automorphisms also contains the group of conformal transformations \cite{Scholz1}.

In the field equations of the Weyl geometric black hole solution, due to the effects of the Weyl geometry, some extra terms, induced by the presence of the scalar field, do appear. These new terms can be interpreted as describing the energy density and pressure of an effective fluid located around the black hole. Hence, the Weyl geometric black hole is surrounded by a material cloud, which may be interpreted physically as forming an accretion disk, or acting as an effective dark matter, as suggested in \cite{HT4}. Hence,  black holes in Weyl geometric theory contain scalar hair, which determines the  strength of the gravitational interaction. Due to the presence of the scalar field, an effective gravitational constant is generated in the spacetime of the black hole, which changes with the distance from the center of the black hole, even after the spontaneous conformal symmetry breaking of the vacuum.

In analyzing the thermodynamic properties of the Weyl geometric black hole we have first discussed the existence, and the borders for horizons of this black hole. The equation giving the position of the horizons is given by Eq.~(\ref{eq:delta}), a third order algebraic equation. For the existence of three real roots the condition (\ref{eq:deltaconstraint}) must be satisfied. The positions and the nature of the event horizon are strongly dependent on the solution parameter $\delta$, whose variation in a relatively small range  induces significant changes in the nature and position of the horizons.

The extra terms in the gravitational field equations, generated by the presence of the scalar field, can be interpreted as generating a fluid with negative pressure.  The total mass, expressed by Eq.~(\ref{eq:massG}), contains  the contribution of the ordinary (baryonic) mass $r_g/2$, plus the contribution of the scalar field, which increases due to the presence of the terms linear, quadratic and cubic in $r$. However, dark matter cannot extend beyond the zero pressure region~(if we define ``matter" as having non-negative pressure), which gives for the total mass of dark matter the finite expression (\ref{eq:Mmax}). It is interesting to note that the total mass of the dark matter is dependent on the baryonic mass.

We have also studied the thermodynamic characteristics of the Weyl geometric black hole, and we have obtained the explicit expressions of the basic thermodynamic quantities, as well as their dependence on the parameters of the solution.

While the general expression of the Hawking temperature of the evaporating Weyl geometric black hole remains similar to the ones for black hole in general relativity, since both can be obtained from the metric tensor components, the entropy has a new scalar field factor, in addition to the standard area law.

By appropriately computing the thermodynamic mass, the first law of black hole thermodynamic of Weyl geometry black hole can be established. The heat capacity for Weyl geometric black holes is negative, but smoothly changes to positive values for de Sitter spacetime, suggesting a continuous phase transition between unstable and stable configuration in the presence of a surrounding, even when the spacetime is asymptotically de Sitter.

The luminosity and evaporation time of Weyl black hole show certain similar result as standard general relativistic  black holes up to the Nariai limit.

We would like to point out that our results on the thermodynamics of Weyl geometric gravity black holes have been obtained for a specific, exact solution of the field equations, under some particular assumptions adopted for the behavior of the auxiliary scalar field, and for the Weyl vector. Therefore, as compared to all the possible spectrum of the black hole solutions in Weyl geometric gravity they may be considered as having a qualitative relevance only. But, even within this restricted level of generality, the exact solution of the field equations of Weyl geometric gravity may provide an important insight into the complex physical and thermodynamic behavior of the Weyl geometric black holes, and of the novel and interesting astrophysical and physical features related to them.

As the analysis of the present work has shown, Weyl geometric gravity black holes have more thermodynamic variability, resulting from their basic properties, as compared with the Schwarzschild black holes of general relativity. This can be easily seen from the study of the Hawking temperature, entropy or luminosity of the Weyl geometric black holes. These novel properties do follow from the presence of the vector and scalar degrees of freedom, and they reflect the complex mathematical structure of the theory, described mathematically by very complicated, and strongly nonlinear field equations. Thus, the physical effects associated with the Weyl vector field and the scalar field, two new degrees of freedom of the theory, could also lead to a new understanding of the laws of black hole thermodynamics, as well as to complex interplay between thermodynamics, gravity, and quantum mechanics. The possible quantum implications of the existence of the Weyl geometric black holes will be considered in a future study.

\section*{Acknowledgments}

M. F. A. R. S. is supported by the Second Century Fund (C2F), Chulalongkorn University, Thailand. PB is supported in part by National Research Council of Thailand~(NRCT) and Chulalongkorn University under Grant N42A660500. This research has received funding support from the NSRF via the Program Management Unit for Human Resources \& Institutional Development, Research and Innovation~[grant number B39G660025]. The work of T.H. is supported by a grant from the Romanian Ministry of Education and Research,
CNCS-UEFISCDI, project number PN-III-P4-ID-PCE2020-2255 (PNCDI III).

\appendix

\section{Basics of Weyl geometry and Weyl geometric gravity}\label{app:1}

We define in a general sense Weyl conformal geometry as the equivalence classes of ($g_{\mu\nu}, \omega_\mu$) of the metric
and of the Weyl gauge field ($\omega_\mu$), related by the Weyl gauge transformations
\bea\label{WGS2}
\hat g_{\mu\nu}\!=\!\Sigma^d 
\,g_{\mu\nu},
\sqrt{\hat{-g}}=\!\Sigma^{2 d} \sqrt{-g}, \hat\omega_\mu\!=\!\omega _\mu -\frac{1}{\alpha}\, \partial_\mu\ln\Sigma,\nonumber\\
\eea
where by $d$ we have denoted the Weyl weight (charge) of  $g_{\mu\nu}$,  while $\alpha$ is the Weyl gauge coupling.
 In our present approach we consider only the case $d=1$, but our results can be generalized to arbitrary $d$ by rescaling the coupling as  $\alpha\rightarrow  \alpha\,\times d$.

The Weyl gauge field is included in the Weyl  connection $\tilde\Gamma$, which  is obtained as the solution of the system
 \bea\label{nablag}
\tilde\nabla_\lambda  g_{\mu\nu}=- d\, \alpha\, \omega _\lambda g_{\mu\nu}
\eea
with $\tilde\nabla_\mu$ defined with the help of $\tilde\Gamma_{\mu\nu}^\lambda$ according to
\bea\label{eeee}
\tilde\nabla_\lambda g_{\mu\nu}=\partial_\lambda g_{\mu\nu}- \tilde \Gamma^\rho_{\mu\lambda}
g_{\rho\nu} -\tilde\Gamma^\rho_{\nu\lambda}\,g_{\rho\mu}.
\eea

From Eq.(\ref{nablag}) it follows that Weyl geometry is {\it non-metric}. Still the covariant derivative can be written as $(\tilde \nabla_\lambda+d\,\alpha\,\omega_\lambda)\,g_{\mu\nu}=0$, a form similar to the Riemannian case. Hence, in Weyl geometry one can use the Riemannian expressions in which
the covariant derivatives $\nabla_\lambda$,  acting on the geometric and physical quantities are replaced by their Weyl-geometric counterparts,
as, for example,  in
$
\partial_\lambda \rightarrow \partial_\lambda+ \rm{weight}\times \alpha \times \omega _\lambda
$,
where 'weight' denotes  the Weyl charge.

The expression of $\tilde \Gamma$ from  Eq.~(\ref{nablag}) is obtained by using the cyclic
permutations of the indices, and combining the resulting equations. Thus we find
\bea\label{AGamma}
\tilde \Gamma_{\mu\nu}^\lambda=
\Gamma_{\mu\nu}^\lambda+\alpha \,\frac{d}{2} \,\Big[\delta_\mu^\lambda\,\omega _\nu
+\delta_\nu^\lambda\, \omega _\mu - g_{\mu\nu} \,\omega ^\lambda\Big],
\eea
where by $\Gamma_{\mu\nu}^\lambda$ we have denoted the Levi-Civita connection of the Riemann geometry
\bea\label{trace}
\Gamma_{\mu\nu}^\lambda=\frac12\,g^{\lambda\rho} (\partial_\mu g_{\rho\nu}+
\partial_\nu g_{\rho\mu}-\partial_{\rho} g_{\mu\nu}).
\eea

It is important to note that $\tilde \Gamma$  is invariant under the group of conformal transformations (\ref{WGS2}). Taking the trace of Eq.~(\ref{AGamma}), and denoting $\Gamma_\mu\equiv \Gamma_{\mu\lambda}^\lambda$ and
 $\tilde\Gamma_\mu\equiv \tilde\Gamma_{\mu\lambda}^\lambda$, respectively, we obtain
\bea
\tilde \Gamma_\mu=\Gamma_\mu + 2 d \,\alpha\,\omega _\mu.
\eea

Hence, the Weyl gauge field can be interpreted as describing the deviation
of  the Weyl connection from the Levi-Civita connection. One can obtain the scalar and the tensor curvatures of the Weyl geometry,
using expressions similar to those in Riemannian geometry,  but with $\tilde\Gamma$ substituting $\Gamma$. Thus
\bea
\tilde R^\lambda_{\mu\nu\sigma}&=&
\partial_\nu \tilde\Gamma^\lambda_{\mu\sigma}
-\partial_\sigma \tilde\Gamma^\lambda_{\mu\nu}
+\tilde\Gamma^\lambda_{\nu\rho}\,\tilde\Gamma_{\mu\sigma}^\rho
-\tilde\Gamma_{\sigma\rho}^\lambda\,\tilde\Gamma_{\mu\nu}^\rho,\nonumber\\
\tilde R_{\mu\nu}&=&\tilde R^\lambda_{\mu\lambda\sigma},\nonumber\\
\tilde R&=&g^{\mu\sigma}\,\tilde R_{\mu\sigma}.
\eea

After some simple calculations one obtains
\bea\label{tRmunu}
\tilde R_{\mu\nu}&=&R_{\mu\nu}
+\frac 12\, (\alpha d) (\nabla_\mu \omega_\nu-3\,\nabla_\nu \omega_\mu - g_{\mu\nu}\,\nabla_\lambda \omega ^\lambda)\nonumber\\
&&+\frac12 \,(\alpha d)^2\, (\omega_\mu \omega_\nu -g_{\mu\nu}\,\omega_\lambda \omega ^\lambda),\nonumber\\
\tilde R_{\mu\nu}&-&\tilde R_{\nu\mu}=2 \,d\alpha \, F_{\mu\nu},\nonumber\\
\tilde R&=& R-3 \,d\,\alpha\,\nabla_\mu\omega ^\mu-\frac{3}{2}\,(d\,\alpha )^2 \,\omega _\mu\,\omega^\mu.
\eea
In the above equations the right hand side  is defined in the Riemannian geometry, and hence $\nabla_\mu$ is defined via
the Levi-Civita connection ($\Gamma$).

An important property of Weyl geometry is that $\tilde R$  transforms covariantly under the transformations (\ref{WGS2})
\bea
\hat{\tilde R}=(1/\Sigma^d)\,\tilde R,
\eea
a result which follows from the transformation rule of $g^{\mu\sigma}$,  and from the property of invariance of $\tilde R_{\mu\nu}$ under the conformal transformations, since
$\tilde \Gamma$ is also invariant. Then, it follows immediately that the term $\sqrt{g}\, \tilde R^2$
is also Weyl gauge invariant.

An important geometrical quantity is the strength of the Weyl vector field $\tilde{F}_{\mu \nu}$, defined according to
\be
\tilde{F}_{\mu \nu}=\nabla _\mu \omega _\nu-\nabla _\nu \omega _\mu.
\ee

The simplest conformally invariant gravitational Lagrangian density is given by
\begin{equation}
L_{W}=\left( \frac{1}{4!\,\xi ^{2}}\tilde{R}^{2}-\frac{1}{4}\tilde{F}_{\mu
\nu }\tilde{F}^{\mu \nu }\right) ,  \label{inA}
\end{equation}%
where  $\xi <1$ is the parameter of the perturbative
coupling.

The Lagrangian $L_{W}$ can be linearized with the help of the substitution
\begin{equation}
\tilde{R}^{2}\rightarrow -2\phi ^{2}\,\tilde{R}-\phi ^{4},
\end{equation}%
where $\phi $ is an auxiliary scalar field. Then the Lagrangian becomes
\begin{equation}
L_{W}=\left( -\frac{\phi ^{2}}{12\xi ^{2}}\tilde{R}-\frac{\phi ^{4}}{%
4!\,\xi ^{2}}-\frac{1}{4}\tilde{F}_{\mu \nu }\tilde{F}^{\mu \nu }\right)
\sqrt{-\tilde{g}}.  \label{alt3}
\end{equation}%
Substituting into Eq.~(\ref{alt3}) the expression of $\tilde{R}$,
\begin{equation}
\tilde{R}=R-3\alpha \nabla _{\mu }\omega ^{\mu }-\frac{3}{2}\alpha
^{2}\omega _{\mu }\omega ^{\mu },  \label{R}
\end{equation}%
we obtain the action of the Weyl geometric gravitational theory as given by
\begin{eqnarray}
\mathcal{S} &=&\int \Bigg[-\frac{\phi ^{2}}{12\xi ^{2}}\Big(R-3\alpha
\nabla _{\mu }\omega ^{\mu }-\frac{3}{2}\alpha ^{2}\omega _{\mu }\omega
^{\mu }\Big)  \notag \\
&&-\frac{\phi ^{4}}{4!\,\xi ^{2}}-\frac{1}{4}\tilde{F}_{\mu \nu }\tilde{F%
}^{\mu \nu }\Bigg]\sqrt{-g}\,d^{4}x,
\end{eqnarray}

\end{document}